\documentclass[alpha-refs]{oup-contemporary}

\usepackage{graphicx}
\usepackage{siunitx}
\usepackage{amsmath}
\usepackage{longtable,tabularx}
\usepackage{algorithm}

\journal{jcde}
\title{Airfoil GAN: Encoding and Synthesizing Airfoils for Aerodynamic Shape Optimization}
\doi{10.1093/jcde/xxxxxxx}
\jyear{202X}
\jvolume{X}
\jnumber{X}
\jstartpage{X}
\jendpage{X}
\jpubdate{Day Month Year}

\author[1]{Yuyang Wang}
\author[1]{Kenji Shimada}
\author[1]{Amir Barati Farimani}

\affil[1]{Department of Mechanical Engineering, Carnegie Mellon University, Pittsburgh, PA 15213, USA}

\authnote{\authfn{1}Corresponding author. E-mail: \href{mailto: abc@uni.edu}{barati@cmu.edu}}

\papercat{Research article}

\runningtitle{Airfoil GAN: Encoding and Synthesizing Airfoils for Aerodynamic Shape Optimization}

\begin{document}

\begin{frontmatter}{a}{b}{c}
\maketitle
\begin{abstract}
The current design of aerodynamic shapes, like airfoils, involves computationally intensive simulations to explore the possible design space. Usually, such design relies on the prior definition of design parameters and places restrictions on synthesizing novel shapes. 
In this work, we propose a data-driven shape encoding and generating method, which automatically learns representations from existing airfoils and uses the learned representations to generate new airfoils. The representations are then used in the optimization of synthesized airfoil shapes based on their aerodynamic performance. Our model is built upon VAEGAN, a neural network that combines Variational Autoencoder with Generative Adversarial Network and is trained by the gradient-based technique. Our model can (1) encode the existing airfoil into a latent vector and reconstruct the airfoil from that, (2) generate novel airfoils by randomly sampling the latent vectors and mapping the vectors to the airfoil coordinate domain, and (3) synthesize airfoils with desired aerodynamic properties by optimizing learned features via a genetic algorithm. Our experiments show that the learned features encode shape information thoroughly and comprehensively without predefined design parameters. By interpolating/extrapolating feature vectors or sampling from Gaussian noises, the model can automatically synthesize novel airfoil shapes, some of which possess competitive or even better aerodynamic properties comparing to airfoils used for model training purposes. By optimizing shapes on the learned latent domain via a genetic algorithm, synthesized airfoils can evolve to target aerodynamic properties. This demonstrates an efficient learning-based airfoil design framework, which encodes and optimizes the airfoil on the latent domain and synthesizes promising airfoil candidates for required aerodynamic performance. 
\end{abstract}

\begin{keywords}
Deep Learning; Generative Adversarial Network; Airfoil Design; Design Optimization
\end{keywords}
\end{frontmatter}


\setcounter{section}{0}


\section{Introduction}
\noindent
Geometry parameterization plays an important role in shape design (\cite{samareh2001survey,chang20113d,salunke2014airfoil,masters2017geometric}), and geometry heavily influences the performance, especially in the design of aerodynamic products like airfoils. A practical and effective airfoil design must meet certain aerodynamic requirements, like lift, drag, pitching moment, and critical-speed characteristics (\cite{abbott1945summary}). Due to such multimodality of the airfoil design space, gradient-free methods is commonly applied to optimize the airfoil design (\cite{rajnarayan2008multifidelity}). For feasible and efficient optimization, various airfoil shape parameterization or dimension-reduction methods are implemented to define a lower-dimensional design space. The selection of the design space greatly influences the performance of the airfoil design optimization. Traditional parameterization or dimension-reduction techniques rely on manually selected design parameters like control points of B\'ezier curves (\cite{sederberg1992bezier}) or B-splines (\cite{derksen2010bezierparsec}), which places restrictions on the generalization to various shapes as well as synthesizing novel geometries. 

Recent years have witnessed the success of deep learning (\cite{lecun2015deep}) in many fields like computer vision (\cite{voulodimos2018deep}), natural language process (\cite{cambria2014jumping}) and robotics (\cite{pierson2017deeprobot,carrio2017review}). Such data-driven methods can automatically learn compact and comprehensive representations from samples. However, most of the prevalent deep learning models are based on supervised learning, meaning the samples are paired with manually tagged labels. Such supervision makes the model hard to generalize since both the amount of labeled data and the information contained in the label are limited. Hence, self-supervised learning is proposed to learn the features directly from data. Autoencoder (AE) (\cite{kramer1991nonlinear}) proposes to learn latent features through an encoder-decoder architecture, where the encoder down-samples high-dimensional input into a latent feature domain while the decoder reconstructs the sample from learned low-dimensional feature. By minimizing the difference between the reconstructed sample and the original one, AE automatically learns features without any labels. However, AE performs poorly in generating novel samples from the latent domain. To this end, Variational Autoencoder (VAE) (\cite{kingma2013auto,rezende2014stochastic}) introduces the KL divergence regularization on the latent feature domain with respect to a prior standard normal distribution while follows the insight of the encoder-decoder. Therefore, VAE can generate novel samples by decoding the latent vectors sampled from the prior distribution. Generative Adversarial Network (GAN) (\cite{goodfellow2014gan}) pushes learning from self-supervision even further via a min-max game between a generator and a discriminator. The discriminator works as a classifier to determine real samples from synthesized fake ones. Meanwhile, the generator, which synthesizes samples from random noise, is intended to cheat the discriminator. By jointly training the two components, GAN can generate super high-quality realistic samples (\cite{karras2019style}), which VAE fails to achieve. Through learning to reconstruct or synthesizing samples, the self-supervised models automatically encode high-dimensional input into informative features, which can be generalized to different tasks without restrictions from human labels. Implementation of self-supervised deep learning methods on shape parameterization, like VAE and GAN, can help overcome the limitations of traditional techniques and synthesize shapes with great novelty, which can provide insights for future geometric design. 

In this work, we intend to use deep generative models for the following challenges in shape designs: (1) learning of expressive representations that model the design, (2) generation of realistic designs, and (3) optimization of generated designs toward desired targets. In particular, our work focuses the airfoil design. To this end, we adapt the insight of VAEGAN (\cite{larsen2015vaegan}) and curate the architecture and losses. VAEGAN takes advantage of both VAE and GAN. With the encoder-decoder architecture from VAE, the model learns to explicitly encode an existing airfoil shape into a low-dimensional feature domain and reconstruct the shape with little error. Also, the KL divergence regularizes the latent features to follow the prior standard normal distribution, so that latent vectors can be sampled from the prior distribution. With the discriminator from GAN, our decoder automatically learns to generate a large number of high-quality novel airfoils from the prior distribution by adversarially competing with the discriminator. Moreover, we introduce the genetic algorithm to VAEGAN generated airfoil designs to optimize the aerodynamic properties (i.e., lift and drag coefficients). We further apply the genetic algorithm to optimize the generated airfoils from VAEGAN toward desired aerodynamic properties. 

Our experiments show that generated airfoils are smooth even without any smoothing post-process. K-means clustering in the learned feature domain demonstrates that feature vectors encode essential shape information in a way that each cluster represents various shape patterns. Further test on learned latent features illustrates that different geometry information is encoded in each dimension of the representation. The performance of the model in synthesizing novel airfoils is examined as well. By either interpolation or extrapolation of feature vectors, a synthesized airfoil inherits features of parent samples, while generated airfoils from sampled Gaussian noise show great novelty in a way that it is not a simple combination of two existing airfoils. A further experiment on the aerodynamic properties of synthesized airfoils, either by interpolating, extrapolating, or sampling, indicates the synthesized airfoils can possess competitive aerodynamic properties, and some even surpass the existing ones. With a genetic algorithm (\cite{mitchell1998introduction}), airfoil geometries can be optimized on the feature domain and evolve to possess specific aerodynamic properties. Our model proves its ability to parameterize the existing airfoil shape as well as generating novel and practical airfoils; both lead to designing the new generation of airfoils more intelligently and efficiently without intensely relying on experimental experience or manually selected design parameters. Overall, experiments have demonstrated that the deep generative model presented in our work learns meaningful shape representations and generates realistic airfoil designs. Besides, the generated designs can be optimized toward target properties when directly combined with optimization methods. 


\section{Related Works}
\noindent
We propose to use a generative model to synthesize novel airfoils without any predefined design parameters. Through training via the gradient-based method, our deep learning model also automatically learns to parameterize airfoils into latent feature vectors. To better address the insight of our work, this section reviews previous work on shape parameterization, especially on aerodynamic geometries and the implementation of self-supervised deep learning on geometry design and synthesis.

\subsection{Geometry Parameterization and Dimension Reduction}
\noindent
A lot of work has been done in parameterizing complex shapes and reducing geometric dimensions, including discrete approach, analytical approach, polynomial and spline approaches, free-form deformation (FFD) approach, etc. The discrete approach leverages the grid-point coordinates as design parameters and different shapes are generating by moving individual grid points (\cite{campbell1992approach, jameson1998optimum}). Though easy to implement, such method is likely to synthesize shapes that are not smooth, which can be a challenge for CFD solvers (\cite{jameson2006advances}). Therefore, it is commonly preferred to use more robust and efficient shape parameterization strategies rather than the discrete approach. 

In analytical approach, a compact formulation for parameterization is obtained by adding analytic shape functions to the baseline shape (\cite{hicks1978wing}). The design variables, in this case, are the coefficients correlated with shape functions. Also, \cite{hager1992multi} follows the same formulation strategy but with different shape functions. The such method works well for several families of airfoils, but it can fail in representing radical new airfoil designs. 

Polynomial and spline are also utilized in shape parameterization. With different orders of polynomials as the basis, the airfoil shape can be described as a linear combination of the basis (\cite{elliott1997practical,taylor1991sensitivity}). However, high-order terms can overfit to high-frequency noise, especially when coefficients are of different magnitudes. Besides polynomial, B\'ezier curve (\cite{sederberg1992bezier}), which is built upon the Bernstein polynomials, is another mathematical formulation of curves. In detail, $n+1$ control points of B\'ezier are needed to define an $n$-degree B\'ezier curve. Although B\'ezier and polynomial curves are mathematically equivalent, B\'ezier usually perform better in controlling a curve since control points are closely related to the curve position and shape. To mitigate the rounding error, De Casteljau (\cite{boehm1999casteljau}), a recursive algorithm, is introduced to compute the Bernstein polynomials numerically. The B-spline curve with B-spline basis functions is also utilized to describe the airfoil shape. Non-uniform rational B-spline (NURBS) is introduced (\cite{farin2014curves}) to represent both standard geometric objects like lines, circles, ellipses, and cones, as well as free‑form geometries. The polynomial and spline approaches are suited for two dimensional shapes. However, when modeling complex shapes, a large number of control points are need, which can lead to irregular and wavy shapes in optimization. 

FFD utilizes high-level shape deformation instead of lower-level geometric entities to represent a shape. Based on this insight, \cite{sederberg1986free} presents a technique which can apply deformation either locally or globally based on trivariate Bernstein polynomials. The deformation is manipulated by control points of trivariate B\'ezier volumes. In \cite{coquillart1990extended}, an extended free-form deformation (EFFD) method is presented, which enables arbitrarily shaped deformations by using non-parallelepiped lattices. Research presented in \cite{yeh1998applying} incorporates FFD and sensitivity analysis where geometry changes and structural responses are correlated, and the shape satisfying deformation or stress constraints can be found easily. 

Other physically intuitive approaches, like PARSEC (\cite{sobieczky1997parsec}), defines geometric parameters to express the airfoil shape, including upper and lower curvature, thickness, leading-edge radius, etc. B\'ezier-PARSEC (\cite{derksen2010bezierparsec}), which combines B\'ezier and PARSEC, uses PARSEC parameters to define B\'ezier curves. Besides, camber and thickness mode shapes from the existing airfoils are utilized to parameterize airfoil shapes (\cite{li2019data}). Also, the linear reduction method like the SVD is utilized to extract airfoil representations and optimize shape design (\cite{poole2019efficient}). Parametric model embedding \cite{serani2023parametric} extends the formulation of Karhunen–Loève expansion \cite{diez2015design} via a generalized feature space that includes design variables vectors and a generalized inner product to select the latent design space. Active subspace \cite{constantine2014active} leverages gradient to detect the directions of the strongest variability and uses the directions as low-dimensional active subspace to parameterize surfaces. \cite{ghorbani2021airfoil} extends B-spline to reconstruct airfoil shapes from noisy 3D point clouds. The conventional dimension reduction or parameterization techniques have been implemented in different scenarios and successfully represented the existing airfoil shapes. However, these methods usually require pre-defining the design space as well as the boundary of design space, like the design parameters in PARSEC, shape functions, etc., which can degrade the synthesis of novel/new airfoils and optimization towards the desired design.


\subsection{Deep Learning in Geometry Design}
\noindent
In recent years, deep learning has been successful at extracting informative features from data (\cite{lecun2015deep}). Supervised learning, which maps an input to an output based on training input-output pairs, has become general solutions in many fields, like image classification (\cite{lecun1989lenet,szegedy2015going,he2015delving}), object detection (\cite{ren2015fasterrcnn,redmon2016yolo}), and segmentation (\cite{li2017fully,he2017maskrcnn}). 
However, supervised learning is limited to labeled data, and collection of a large number of labels can be expensive. Self-supervised learning proposes to utilize the data itself as labels and learn features by predicting the data, which requires no explicitly labeled data. Following the insight, variational Autoencoder (VAE) (\cite{kingma2013auto}), learns representative latent features by reconstructing the input. 
Also, with the introduction of GAN (\cite{goodfellow2014gan}), deep learning in a self-supervised manner has been widely used in synthesizing realistic samples, and some sophisticated GAN-based model can generate high-resolution images which are even hard to distinguish by humans (\cite{karras2019style}). Due to the ability to extract representative features and synthesize realistic samples, deep learning has been widely used to make the geometric design more systematic and efficient. This section will introduce some work of deep learning in geometric design, especially in aerodynamic shape design. 

In the work of \cite{norgaard1997neural}, a multi-layer perceptron (MLP), which takes the angle of attack and flap setting as input, is trained to predict multiple aerodynamic coefficients, including lift coefficient, drag coefficient, and moment of inertia. Further, \cite{zhang2018application} utilizes CNN with airfoil images as input to learn the lift coefficients of different airfoils in multiple flow Mach numbers, Reynolds numbers, and diverse angles of attack. Similarly, in \cite{yilmaz2017convolutional}, CNN is implemented to predict the pressure coefficient value at the test point. Moreover, \cite{yilmaz2018deep} presents a CNN-based method to learn the correlation between airfoil geometry and pressure distribution. The model can also conduct an inverse airfoil design given the pressure. 

On the other hand, deep generative models have been successfully implemented to generate realistic samples. VAE \cite{kingma2013auto} is proposed to generate plausible samples via an encoder-decoder architecture while optimizing the variational lower bound of the log-likelihood of the training data. However, VAE suffers from generating samples that are blurry and averaged over the training samples \cite{cai2019multi}. Besides, GAN, with a discriminator and a generator, is designed to learn features and generate samples without manually tagged labels. The discriminator is a classifier telling the true input from the synthesized fake one, while the generator is intended to generate plausible samples to cheat the discriminator. Since first introduced (\cite{goodfellow2014gan}), many pieces of research have been dedicated to pushing the edge of GAN. In Wasserstein GAN (\cite{arjovsky2017wasserstein}), by introducing Wasserstein distance, the quality of generated samples can be well measured during training. ConditionalGAN (\cite{mirza2014conditional}) and InfoGAN (\cite{chen2016infogan}) extend GAN to generate a sample from various categories within one model. Also, DCGAN (\cite{radford2015unsupervised}) and StyleGAN (\cite{karras2019style}) can generate high-quality realistic samples with sophisticated architectures. 

Deep generative models have been applied to geometry design. D'Agostino et al. \cite{d2018deep} introduce autoencoder to learn a reduced design space and apply optimization on the learned latent space. With the ability to generate plausible samples, GAN provides a powerful architecture to learn representations that can help shape design \cite{nie2021topologygan}. Based on this insight, \cite{chen2019aerodynamic,chen2020airfoil} introduce B\'ezierGAN, which uses a GAN to generate B\'ezier curve control points and then uses the control points to formulate the boundary of airfoils. Such a pipeline guarantees that generated airfoils are smooth, and further shape optimization can be conducted on the feature domain. BSplineGAN (\cite{du2020b}) further extends the insights to B-splines parameterization and proposes a surrogate neural network model to predict lift and drag coefficients. Such methods can be data efficient as they leverage pre-defined formulations of the curves or surfaces. \cite{grey2023separable} propose a data-driven approach that decouples affine and undulation deformations based on Grassmannian-based representation.

Furthermore, the adoption of deep learning-based methods presents significant advantages when dealing with high-dimensional designs. Unlike traditional methods that often struggle to effectively explore such complex design spaces, our approach efficiently navigates the vast parameter spaces associated with advanced and intricate airfoil designs. This ensures that our method can tackle challenging design problems, offering valuable insights and facilitating the creation of cutting-edge designs in a variety of engineering applications

By not limiting ourselves to a specific design space, our VAEGAN method excels at directly capturing the underlying distribution of airfoil shapes from datasets. It achieves this by effectively learning intricate non-linear relationships between the latent features it acquires and the corresponding airfoil designs. An additional advantage is the ability of our method to generate novel airfoil shapes that do not exist in the training dataset. This facilitates the exploration of new design possibilities beyond the constraints of the chosen design space. Furthermore, deep learning-based methods offer efficient solutions for high-dimensional designs, whereas traditional methods often struggle to explore such design spaces effectively.

Conditional GAN has also been utilized to generate airfoil shapes given certain airfoil properties (\cite{achour2020development, yilmaz2020conditional}). Further, GAN can be implemented to generate three-dimensional samples, as shown in \cite{wu20163dgan,huang2015analysis,sinha2017surfnet}. \cite{wu20163dgan} proposes to use a three-dimensional convolutional layer to generate volumetric objects. While \cite{huang2015analysis} trains a generative model of three-dimensional shape surfaces, which directly encodes surface geometry and shape structure, \cite{kalogerakis2012probabilistic} further proposes a model to represent probabilistic relationships between properties of shape components and relates them to learned underlying causes of structural variability within the domain. 

In our work, we propose to use a VAEGAN-based model to extract features of airfoil shapes and synthesize new airfoil designs. It is demonstrated that such a self-supervised learning framework learns representative geometric features without pre-defined design parameters. The learned latent features can be utilized to synthesize new designs through interpolation or extrapolation. In comparison to conventional parameterization methods, our model can generate a wider variety of new airfoils, and some show promising aerodynamic properties. By applying the genetic algorithm, the VAEGAN synthesized airfoils can be optimized to desired aerodynamic properties. Unlike previous works, like B\'ezier GAN and B-Spline GAN, that rely on pre-defined design formulations like B\'ezier or B-Spline curves, we do not leverage prior formulation of the design and apply deep neural networks as universal function approximators to model a wide range of designs. By not imposing a specific design space, our proposed VAEGAN method is able to capture the underlying distribution of airfoil shapes directly from datasets more effectively by learning complex non-linear relationships between the learned latent feature and airfoil designs. Moreover, such a method is expected to generate novel airfoil shapes that are not present in the training dataset. This facilitates the exploration of new design possibilities beyond the constraints of the chosen design space. Besides, such deep learning-based methods can be extended to high-dimensional designs, where well-defined formulations are hard to comprehensively model the objects. Using the learned latent feature, optimization can navigate designs toward desired properties. 


\section{Method}

\subsection{Data Pre-processing}
\begin{figure}[bt!] 
\centering
    \includegraphics[width=\linewidth]{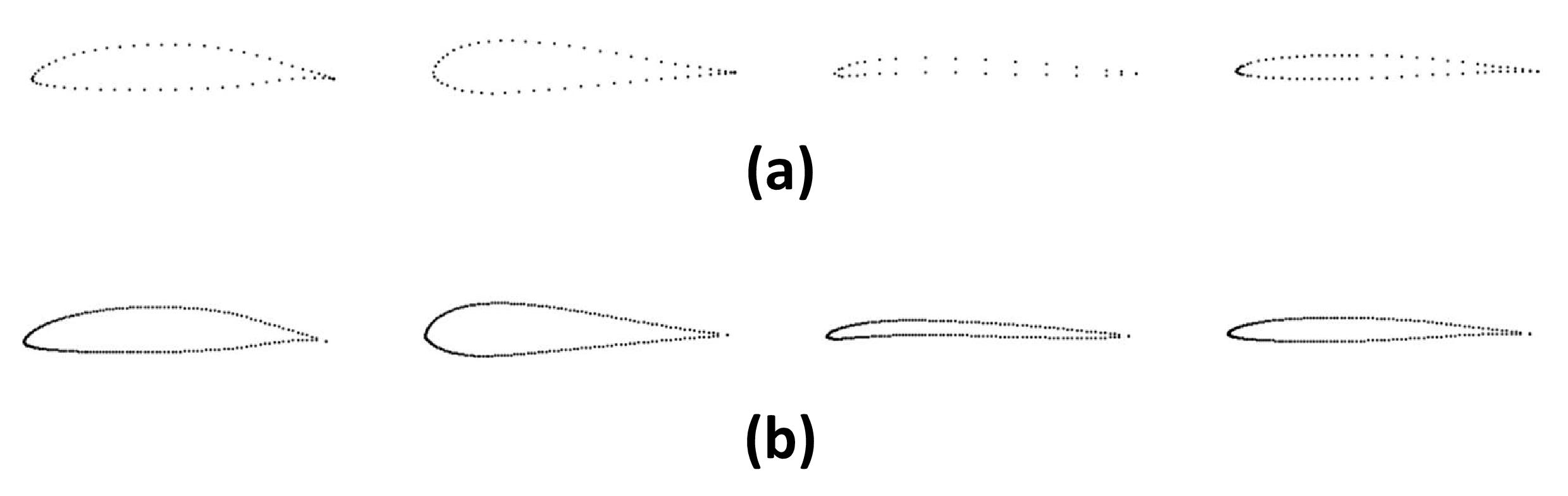}
\caption{(a) Origin airfoil coordinates from the UIUC database; (b) corresponding pre-processed airfoils.}
\label{fig:preprocess}
\end{figure}

\noindent
The UIUC Coordinates Database (\cite{uiucdataset}), which contains more than 1,600 2-dimensional airfoils, is used to train the generative model. The x coordinates of all airfoils from the database have been scaled to $[0,1]$. However, each airfoil in the database is represented by varying numbers of points with different x and y coordinates. Such variation restricts data to be fed directly into a neural network model, which requires a homogeneous input. To deal with it, the upper boundary and the lower boundary are interpolated by cubic splines, respectively. Within each boundary, $N$ points are selected by the x coordinate given in Eq.~\ref{eq:x}.
\begin{equation}
\begin{split}
    \theta_i &= \frac{\pi (i-1)}{N}, \\
    x_i &= 1 - \cos(\theta_i),
\end{split}
\label{eq:x}
\end{equation} 
where $i$ denotes the index of each point, and $N$ is set to 100 in our case.
By this means, all the interpolated airfoils have 200 points in total and share the same x coordinates. Therefore, only the y coordinates of each airfoil are fed into the model, which reduces the dimensionality of the data. Finally, all y coordinates are scaled to $[-1, 1]$ by multiplying a normalization coefficient. As illustrated in Fig~\ref{fig:preprocess}, the first row shows the original airfoils from the UIUC Coordinate Database, while the second row shows the corresponding processed airfoils. 

\subsection{VAEGAN}
\noindent
Our model is based upon VAEGAN (\cite{larsen2015vaegan}), which takes advantage of both VAE (\cite{kingma2013auto, rezende2014stochastic}) and GAN (\cite{goodfellow2014gan}). VAE contains two components: an encoder and a decoder. The former encodes a high-dimensional sample, $x$, into a low-dimensional latent representation, $z$. While the decoder takes as input the latent vector, $z$, and upsamples from the representation domain to the original data domain, $\tilde{x}$. The encoder and decoder are given as:
\begin{equation}
    z \sim \text{Enc}(x) = q(z|x), \: \tilde{x} \sim \text{Dec}(z) = p(\tilde{x}|z),
\label{eq:vae}
\end{equation}
where $q(z|x)$ denotes the distribution of latent vector $z$ given airfoil $x$ and $p(\tilde{x}|z)$ denotes the distribution of reconstructed airfoil $\tilde{x}$ given $z$.
To regularize the encoder, VAE takes into consideration a prior distribution of the latent vector, $p(z)$. Here it is assumed that $z \sim \mathcal{N}(0,\mathbf{I})$, which follows an isotropic Gaussian distribution. The loss function for VAE to minimize is given by:
\begin{equation}
    \mathcal{L}_{VAE} = \mathcal{L}_{recon} + \mathcal{L}_{prior}, 
\label{eq:loss_vae}
\end{equation} with 
\begin{equation}
\begin{aligned}
    \mathcal{L}_{recon} &= ||\tilde{x} - x||_2^2, \text{ and}  \\
    \mathcal{L}_{prior} &= D_{KL}(q(z|x)||p(z)), 
\label{eq:loss_recon_prior}
\end{aligned}
\end{equation} 
where $\mathcal{L}_{recon}$ measures how well the reconstructed data, $\tilde{x}$, is comparing to the original $x$ by the mean square error (MSE), and $\mathcal{L}_{prior}$ is the Kullback Leibler divergence (KL divergence), which measures the difference between encoded representation vectors and Gaussian distribution. 

VAE learns the representation of samples and can reconstruct them from $z$ with the encoder-decoder architecture. However, it suffers from poor performance in generating novel samples, which have not been seen before (\cite{goodfellow2014gan} \cite{larsen2015vaegan}). To this end, a generative adversarial network (GAN)(\cite{goodfellow2014gan}) is introduced, which contains a discriminator, $\mathcal{D}$, and a generator, $\mathcal{G}$, competing with each other in a self-supervised manner. $\mathcal{G}$ tries to generate plausible samples to fool $\mathcal{D}$, while $\mathcal{D}$ keeps sharpening its decision boundary to determine synthesized fake samples from real ones. In detail, the generator, $\mathcal{G}$, is fed with random noise $\hat{z} \sim p(\hat{z})$ and maps the noise to the data sample domain to generate fake $\hat{x}$. The discriminator takes both $x$ and $\hat{x}$ to predict whether the input is from the real dataset or generated by $\mathcal{G}$. The cross-entropy loss function for the min-max game is given as:
\begin{equation}
     \min_G \max_D \mathcal{L}_{GAN} =\log \mathcal{D}(x) + \log(1-\mathcal{D}(\mathcal{G}(\hat{z}))).
\label{eq:loss_gan}
\end{equation}

\begin{figure}[bt!] 
\centering
    \includegraphics[width=\linewidth]{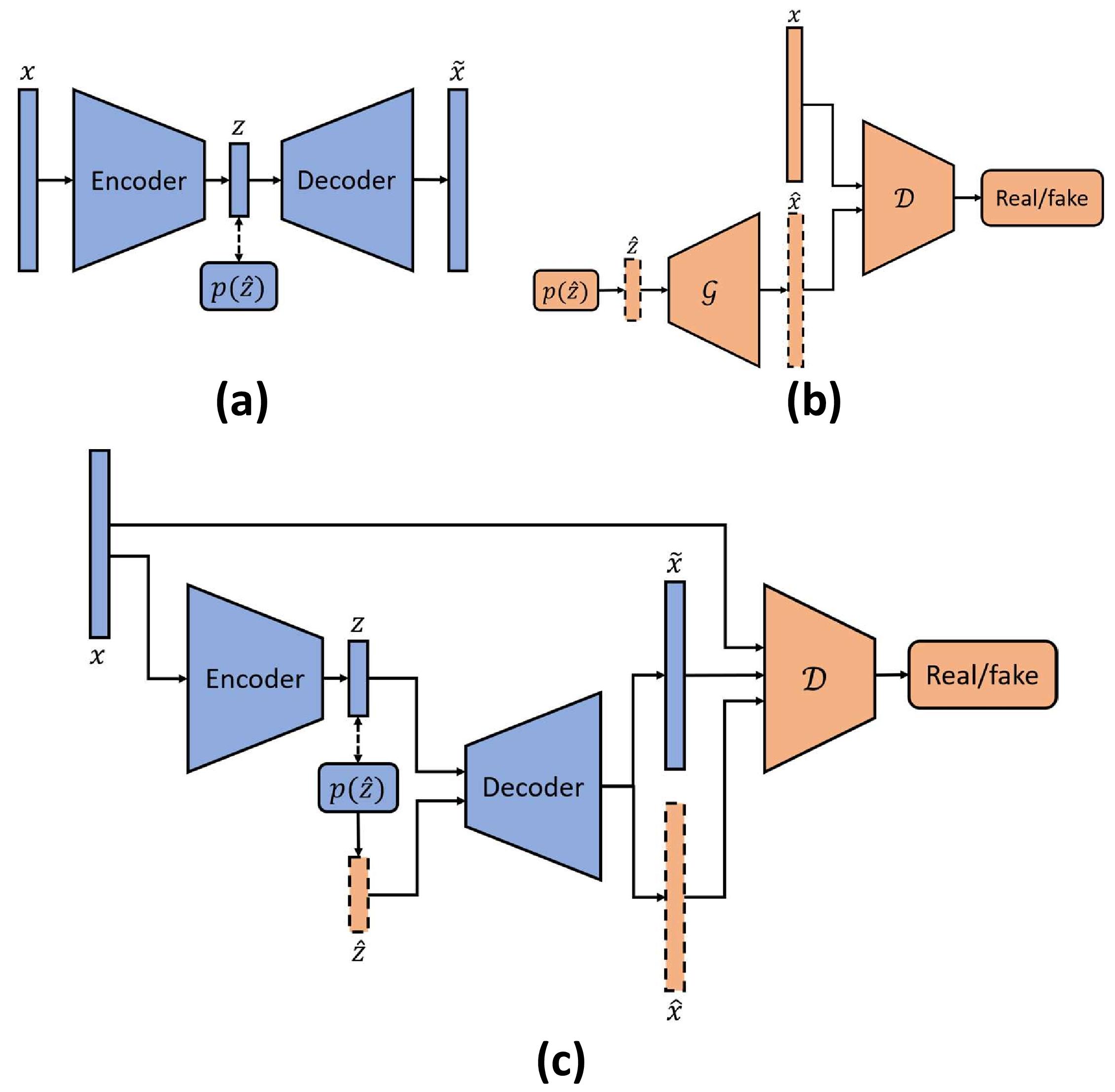}
\caption{Overview of (a) VAE, (b) GAN, and (c) VAEGAN, which integrates the encoder-decoder and discriminator.}
\label{fig:model_overview}
\end{figure}

GAN generates samples from random noise and lacks the mechanism of learning latent representations of the input. Therefore, it is hard to directly learn compact features and optimize the generated airfoils in the latent feature domain using the original GAN. In practice, we not only care about generating new airfoil designs but also expect airfoils with optimized aerodynamic properties. Driven by this, we build our model upon VAEGAN, which combines VAE and GAN. Namely, the generator is replaced by an encoder-decoder structure from VAE, as shown in Figure \ref{fig:model_overview}. Notice that in VAEGAN, the model generates a reconstructed sample, $\tilde{x}$, given a real sample, $x$, and meanwhile generates a fake sample, $\hat{x}$, directly from noise, $\hat{z}$. Both $\tilde{x}$ and $\hat{x}$ should be classified as fake by the discriminator, $\mathcal{D}$, and only $x$ is recognized as the real sample. Hence GAN loss function $\mathcal{L}_{GAN}$ from Eq.~\ref{eq:loss_gan} is modified to:
\begin{equation}
    \mathcal{L}_{GAN} = \log (\mathcal{D}(x)) + \log(1-\mathcal{D}(\text{Dec}(z))) + \log (1 - \mathcal{D}(\text{Dec}(\text{Enc}(x)))),
\end{equation} 
which takes into consideration the real sample, $x$, reconstructed sample, $\tilde{x}$, and fake sample, $\hat{x}$. Besides, to stabilize the training process and sharpen the decision boundary of $\mathcal{D}$, another loss function, $\mathcal{L}_{layer}$, is introduced when training the encoder and decoder. $\mathcal{L}_{layer}$, as given in Eq.~\ref{eq:loss_layer}, measures the $l_1$ distance between the values of the neurons in one particular layer of $\mathcal{D}$ when fake samples are fed and the values when real samples are fed. Such a layer consistency loss works as a supplementary to reconstruction loss which pushes the autoencoder to better reconstruct the real airfoil designs. In our case, we calculate $\mathcal{L}_{layer}$ using the second last hidden layer in the discriminator following Larsen et al \cite{larsen2015vaegan}. 
\begin{equation}
    \mathcal{L}_{layer} = ||\mathcal{D}_l (x) - \mathcal{D}_l (\text{Dec}(\hat{z}))||_1.
\label{eq:loss_layer}
\end{equation} The complete loss function for VAEGAN is a weighted combination of all the loss terms given by:
\begin{equation}
    \mathcal{L} = \lambda_0 \mathcal{L}_{prior} + \lambda_1 \mathcal{L}_{recon} + \lambda_2 \mathcal{L}_{layer} + \lambda_3 \mathcal{L}_{GAN}.
\label{eq:loss_vaegan}
\end{equation} The three components, encoder, decoder and discriminator are trained jointly, and each term of the loss function is assigned with different weights $\lambda$ when training each component. 

\subsection{Airfoil Synthesis}
\noindent
By training on the UIUC Database, the VAEGAN model automatically learns to encode airfoils into latent features and reconstruct airfoils from the feature domain. The learned latent features can be directly utilized for dimension reduction and shape parameterization. Moreover, the VAEGAN model is intended to synthesize novel airfoils which are different from samples in the training dataset. To this end, we propose three synthesis methods: interpolation, extrapolation, and sampling, all of which are conducted on the latent feature domain. 

More specifically, in interpolation or extrapolation, two airfoils from the UIUC Database are first mapped to latent feature vectors, $z_1$ and $z_2$, via a well-trained encoder. A new feature vector $\bar{z}$, which is an affine combination of $z_1$ and $z_2$, is calculated as given in Eq.~\ref{eq:inter_2}: 
\begin{equation}
    \bar{z} = \nu z_1 + (1-\nu) z_2,
\label{eq:inter_2}
\end{equation}
where $\nu$ is the coefficient controlling the weight between $z_1$ and $z_2$. When $0 \leq \nu \leq 1$, $\bar{z}$ is an interpolated feature vector, else it is an extrapolation between $z_1$ and $z_2$. The interpolated/extrapolated feature vector, $\bar{z}$, is then fed into the decoder to synthesize an airfoil. Also, such interpolation/extrapolation between two airfoils can be directly extended to a triplet case. Given $z_1$, $z_2$, and $z_3$ are three feature vectors mapped from three different airfoils via the encoder, the expression of triplet interpolation/extrapolation is shown in Eq.~\ref{eq:inter_3}: 
\begin{equation}
    \bar{z} = \alpha z_1 + \beta z_2 + \gamma z_3, 
    \text{ where } \alpha+\beta+\gamma=1. 
\label{eq:inter_3}
\end{equation}
Similarly, when $0 \leq \alpha,\beta,\gamma \leq 0$, $\bar{z}$ is an interpolation of the three feature vectors, and an extrapolation otherwise. 

Besides interpolation and extrapolation, sampling is another method to synthesize novel airfoils. Unlike interpolation or extrapolation, which relies on feature vectors from existing airfoils, sampling generates airfoils directly from random noise. A feature vector $\hat{z}$ is randomly sampled from an isotropic Gaussian distribution $\mathcal{N}(0,\mathbf{I})$ and then mapped to an airfoil via the decoder. By this means, synthesized airfoils from sampling are less restricted since sampled latent vectors are not constrained by features extracted from airfoils in the UIUC Database and are more likely to introduce novelty to the synthesized shapes. Interpolation, extrapolation, and sampling are widely used methods to generate new samples with deep learning based models. Through interpolation and extrapolation, the model is evaluated whether it learns expressive representations to encode the airfoil designs. While sampling is more flexible to generate different and novel designs. To generate more practical airfoils, optimization can be combined with sampling to manipulate the generation as we introduce in the following section.

\subsection{Aerodynamic-aware Shape Optimization}
\noindent
So far, how the VAEGAN model is built and used to generate novel airfoils has been introduced. However, the novelty in shape does not guarantee a better airfoil design. To design engineering effective airfoils, aerodynamic properties are supposed to be considered. To this end, we propose to use a genetic algorithm (GA) (\cite{poon1995genetic, mitchell1998introduction}) to optimize airfoil shapes by controlling feature vectors learned from the VAEGAN model so that the airfoils can evolve to have the desired aerodynamic properties. Specifically, lift coefficient, $C_l$, and drag coefficient, $C_d$, which measure the aerodynamic force perpendicular and horizontal to the direction of motion, are considered to evaluate the aerodynamic performance of the synthesized airfoils.

As a non-gradient optimization technique, GA is inspired by natural selection and is intended to force individuals to gradually evolve to the optimal. Assume the GA has $N$ generations in total and $M$ individuals in each generation. In our case, individuals are feature vectors. We use $z_i$ to represent all individuals in the $i$\textsuperscript{th} generation, and $z_{i,j}$ for the $j$\textsuperscript{th} individual in the $i$\textsuperscript{th} generation; also the airfoil decoded from $z_{i,j}$ is annotated as $a_{i,j}$. Similarly, $C_l^{i,j}$ and $C_d^{i,j}$ represents lift and drag coefficients of $a_{i,j}$, respectively. The fitness score, $s_{i,j}$, is used to measure the aerodynamic performance of the individual, $z_{i,j}$, as shown in Eq.~\ref{eq:score}: 
\begin{equation}
    s_{i,j} = - (\frac{C_l^{i,j} - C_l^t}{C_l^t})^2 - (\frac{C_d^{i,j} - C_d^t}{C_d^t})^2, 
\label{eq:score}
\end{equation}
where the square of the difference between the target and current aerodynamic coefficients is calculated and normalized by the squared target $C_l^t$ and $C_d^t$. The fitness score is supposed to approach zero as individuals evolve on each generation. As shown in Algorithm~\ref{alg:ga}, the initial generation, $z_0$, is randomly sampled from an isotropic Gaussian distribution, $\mathcal{N}(0,\mathbf{I})$. The GA starts with the selection from the initial generation by randomly picking two individuals and comparing their fitness scores. The one with a higher fitness score wins the tournament and becomes one of the parents. $p_1^i$ and $p_2^i$ denote all the parents 1 and parents 2 in the $i$\textsuperscript{th} generation respectively. Single-point crossover is then implemented to generate offspring from parents 1 and 2. Namely, a crossover point on the parent vector is randomly selected, and all elements after that point are swapped between the two parents. Mutation in the natural selection process is also imitated with additive Gaussian noises. 

\begin{algorithm}
\caption{Aerodynamic Shape Optimization via GA}
\begin{algorithmic}[1]

\Procedure{GA}{$N, M, C_l^t, C_d^t$, $p$} 
    \State Initialize $i := 0$
    \State Sample first generation $z_{0j} \sim \mathcal{N}(0, \mathbf{I})$, for $0<j<M$
    
    \While{$i < N$}
        \State Synthesize airfoils $a_{i,j}$ from $z_{i,j}$ via decoder
        \State Compute $C_l^{i,j}$, $C_d^{i,j}$ and fitness score $s_{i,j} = - (\frac{C_l^{i,j} - C_l^t}{C_l^t})^2 - (\frac{C_d^{i,j} - C_d^t}{C_d^t})^2$ 
        \State Select $M$ parent 1, $p_1^i$, and $M$ parent 2, $p_2^i$, from $z_i$ by tournament
        \State Generate next generation, $z_{i+1}$, through single-point crossover
        \State With probability $p$, $z_{i+1,j}$ will add a Gaussian noise $\mathcal{N}(0,\mathbf{I})$
        \State $i := i+1$
    \EndWhile
    \Return The individual with the highest score from $z_{i}$

\EndProcedure

\end{algorithmic}
\label{alg:ga}
\end{algorithm}


\section{Results \& Discussion}
\noindent
Our VAEGAN model consists of 3 components: an encoder, a decoder, and a discriminator, which are all built on multi-layer perceptron (MLP). We implement the VAEGAN model based on PyTorch framework (\cite{NEURIPS2019_9015}). As illustrated in Fig~\ref{fig:model_overview}, the encoder encodes 200-dimensional airfoil coordinates into a 32-dimensional feature domain while the decoder maps the feature back to the airfoil. 
We select the latent feature dimension, namely the number of design variables, to be 32 as a trade-off between the computational efficiency and the accuracy. Since the reconstruction error of the airfoils barely decreases even with larger latent space dimension.
The discriminator is a classifier examining whether the input is a real airfoil from the UIUC Database, or a fake one reconstructed from the decoder, or synthesized from random noises. In detail, the encoder is modeled by a 3-layer MLP with the number of neurons $[256, 128, 32]$ in each layer, and LeakyReLU (\cite{maas2013leakyrelu}) is implemented as the activation function in each layer. The decoder is also a 3-layer MLP with the number of neurons $[128, 256, 200]$ in each layer. A hyperbolic tangent (Tanh) function works as the activation function in the output layer to scale all outputs into $[-1,1]$. Similarly, the discriminator contains three layers with the number of neurons $[256, 128, 1]$, and outputs the probability of whether the input is real or fake through a Sigmoid activation function. 

To automatically learn the latent features and synthesize airfoils, the VAEGAN model is trained on the UIUC Database for 5,000 epochs, and each epoch goes through all the samples in the dataset. Initial learning rates for all three components: encoder, decoder, and discriminator are set to be $0.0005$ and decay to $0.00005$ after 2500 epochs. The batch size is set to be 16, which is approximately $1/100$ of the database size. Adam optimizer (\cite{kingma2014adam}) is utilized to update all the parameters in the model. As mentioned in Section 3.3, different coefficients are assigned to each term in the loss function Eq.~\ref{eq:loss_vaegan}; also different components, namely the encoder, decoder, and discriminator, have different coefficients, respectively. Coefficients of different loss terms and components are shown in Eq.~\ref{eq:enc_loss}, \ref{eq:dec_loss}, and \ref{eq:d_loss}:
\begin{align}
    \mathcal{L}_{Enc} &= 0.1 \mathcal{L}_{prior} + 0.1 \mathcal{L}_{layer} + 10 \mathcal{L}_{recon}, \label{eq:enc_loss} \\
    \mathcal{L}_{Dec} &= 0.1 \mathcal{L}_{prior} + 0.1 \mathcal{L}_{layer} + 10 \mathcal{L}_{recon} + 5 \mathcal{L}_{GAN}, \label{eq:dec_loss} \\
    \mathcal{L}_{\mathcal{D}} &= \mathcal{L}_{GAN}, \label{eq:d_loss}
\end{align}
where $\mathcal{L}_{Enc}$, $\mathcal{L}_{Dec}$, and $\mathcal{L}_{\mathcal{D}}$ represent loss functions for the encoder, decoder, and discriminator, respectively. A large coefficient was assigned to the reconstruction loss $\mathcal{L}_{recon}$ to ensure the feature contains the information of airfoil geometries. Prior loss $\mathcal{L}_{prior}$ and $\mathcal{L}_{layer}$ are assigned with weight $0.1$ since it can lead to trivial solutions otherwise. The generator considers the adversarial GAN loss $\mathcal{L}_{GAN}$, but instead of minimizing the loss, the generator tried to fool the discriminator by maximizing the term. $\mathcal{L}_{GAN}$ also played an important role in VAEGAN to synthesize realistic samples. Therefore, a large coefficient is assigned to $\mathcal{L}_{GAN}$. 

To better investigate our VAEGAN-based model, we compare the performance with two other paramterization methods, principle component analysis (PCA) and variational autoencoder (VAE) (\cite{kingma2013auto}). PCA conducts a linear transformation from the pre-processed airfoil point coordinates into prioritized latent variables. In our case, the top 32 dimensions are kept as the feature. The VAE follows the same encoder-decoder architecture as the VAEGAN, but lacks the discriminator. The latent feature dimension is also set to 32, and the loss function is given in Eq.~\ref{eq:VAE_loss}:

\begin{equation}
    \mathcal{L}_{Enc} = 0.1 \mathcal{L}_{prior} + 10 \mathcal{L}_{recon}.
\label{eq:VAE_loss}
\end{equation}

\subsection{Airfoil Reconstruction via Encoder-Decoder}

\begin{figure}[bt!] 
\centering
    \includegraphics[width=\linewidth]{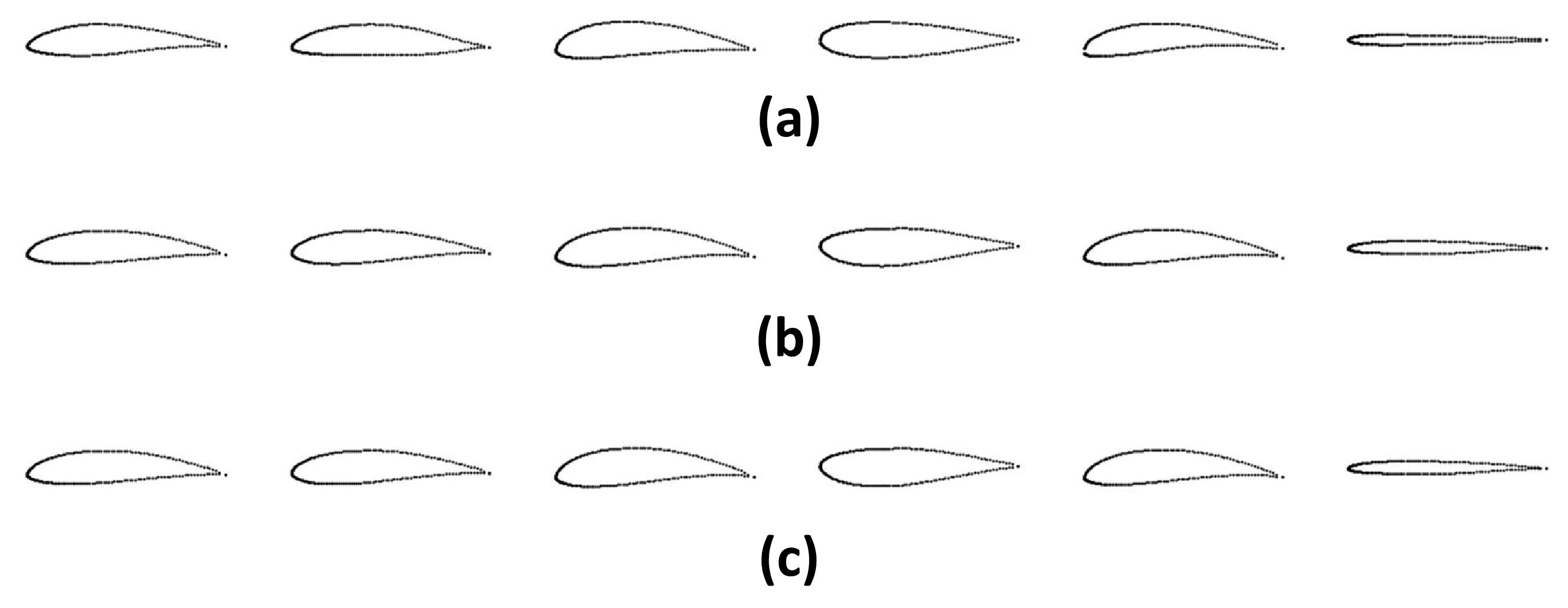}
\caption{(a) Airfoils from UIUC Database, (b) reconstructed airfoils from the VAEGAN, (c) reconstructed airfoils with smoothness.}
\label{fig:recon}
\end{figure}

\begin{figure}[bt!]
\centering
\includegraphics[width=.5\textwidth]{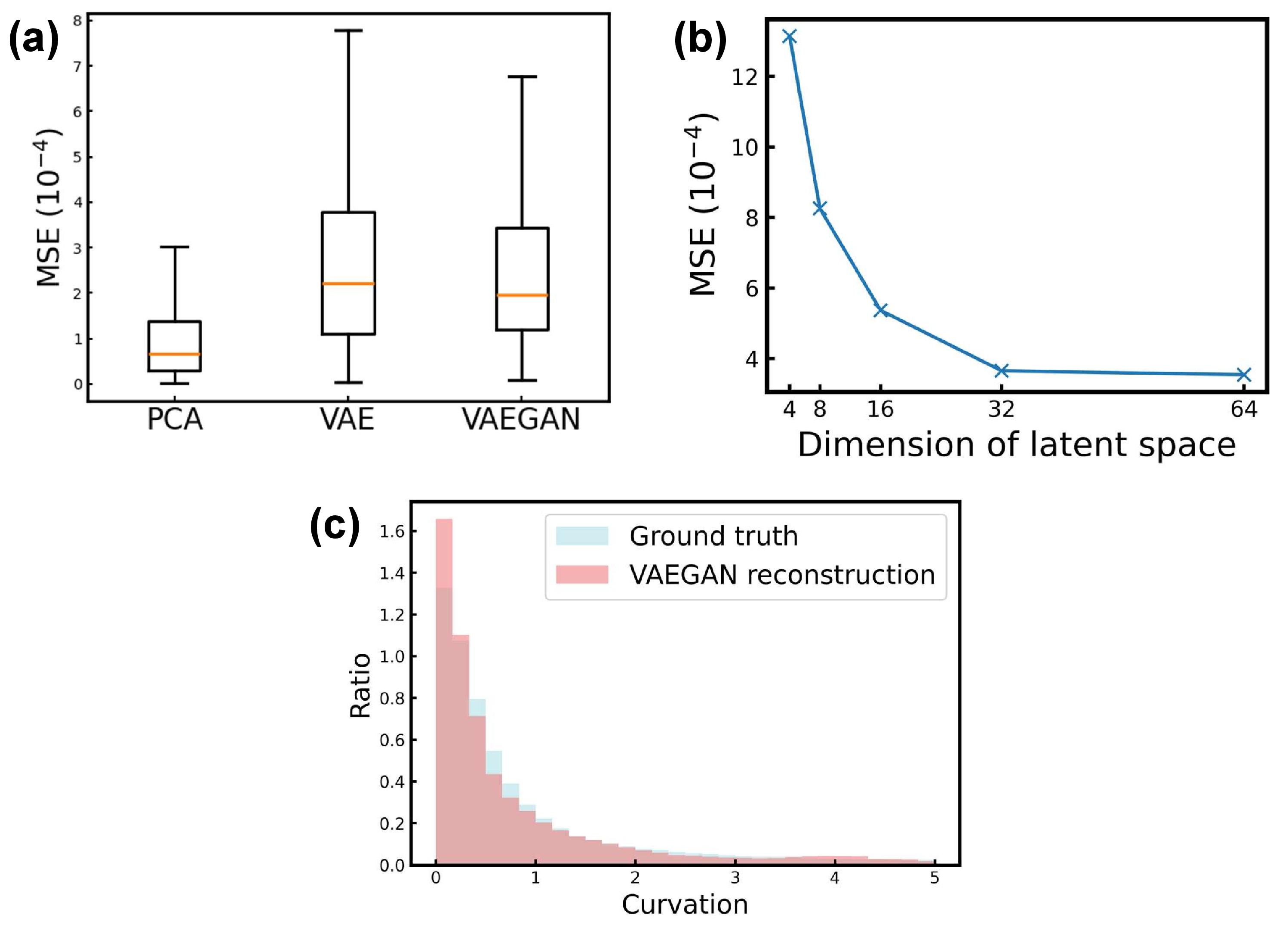}
\caption{(a) Box and whisker plots of MSE error for airfoil reconstruction via different featurization techniques. (b) Reconstruction MSE of VAEGAN with different latent dimensions. (c) Comparison between curvatures of airfoils from dataset and those reconstructed from VAEGAN. }
\label{fig:recon_error}
\end{figure}

\noindent
The VAEGAN model can automatically learn feature vectors, namely mapping the high-dimension airfoils into low-dimension representations. To estimate whether or not the feature vector fully encodes the geometric information of the original airfoil,we first feed airfoils from UIUC Coordinate Database into the encoder to obtain the encoded feature vectors. The decoder then takes the vectors as input and outputs the reconstructed airfoils. Also, the Savitzky-Golay filter (\cite{schafer2011savitzky}), a moving polynomial fitting, is implemented to smoothen the boundary of reconstructed airfoils. In our case, the second-order polynomial is used in the Savitzky-Golay filter, and the length of the moving window is set to 7. In Fig~\ref{fig:recon}, the first row illustrates samples from the UIUC Database, and the second row shows reconstructed airfoils from corresponding feature vectors, with an MSE, 3.65345$\times 10^{-4}$, between the reconstructed and original airfoils. This small error indicates the learned features well represent the shape of airfoils. The third row shows reconstructed airfoils with the Savitzky-Golay filter with an MSE, 3.65054$\times 10^{-4}$, comparing to the original airfoils. These results further demonstrate that the encoder-decoder can reconstruct airfoils that are smooth and realistic without smoothing filters. Also, we compare the reconstruction MSE on test set of our VAEGAN-based model with PCA and VAE as shown in Fig.~\ref{fig:recon_error}a. Such reconstruction error evaluates how models perform in learning expressive representations of the designs \cite{poole2015metric, chen2020airfoil}. our VAEGAN-based model is compared with PCA and VAE as shown in Fig.~\ref{fig:recon_error}a. PCA performs the best in terms of reaching the lowest MSE since it is calculated from a close form solution, whereas VAE and VAEGAN are optimized numerically via the stochastic gradient-based method. With the discriminator and adversarial loss from GAN, VAEGAN model performs slightly better than VAE in reconstruction. It should be pointed out that all three parameterization methods have small reconstruction MSEs of magnitude $10^{-4}$, meaning all the features extracted well encodes the airfoil shapes from the UIUC database. It should also be noted that though similar reconstruction errors are observed across different models, VAEGAN demonstrates preferable performance in generating new airfoil designs as shown in the following sections. We further investigate the effect of latent space dimension on the reconstruction performance as shown in Fig.~\ref{fig:recon_error}b. As the latent dimension increases, the reconstruction MSE decreases. However, when the latent dimension exceeds 32, the improvement of reconstruction quality is trivial. Therefore, we set the latent dimension to 32 as a tradeoff between performance and computational cost. Besides, the curvatures of airfoils reconstructed using VAEGAN are compared with original airfoils as shown in Fig.~\ref{fig:recon_error}c. It is indicated that VAEGAN can generate smooth airfoil designs similar to the ground truth. 


\subsection{Clustering in Feature Domain}
\begin{figure}[bt!]
\centering
\includegraphics[width=0.5\textwidth]{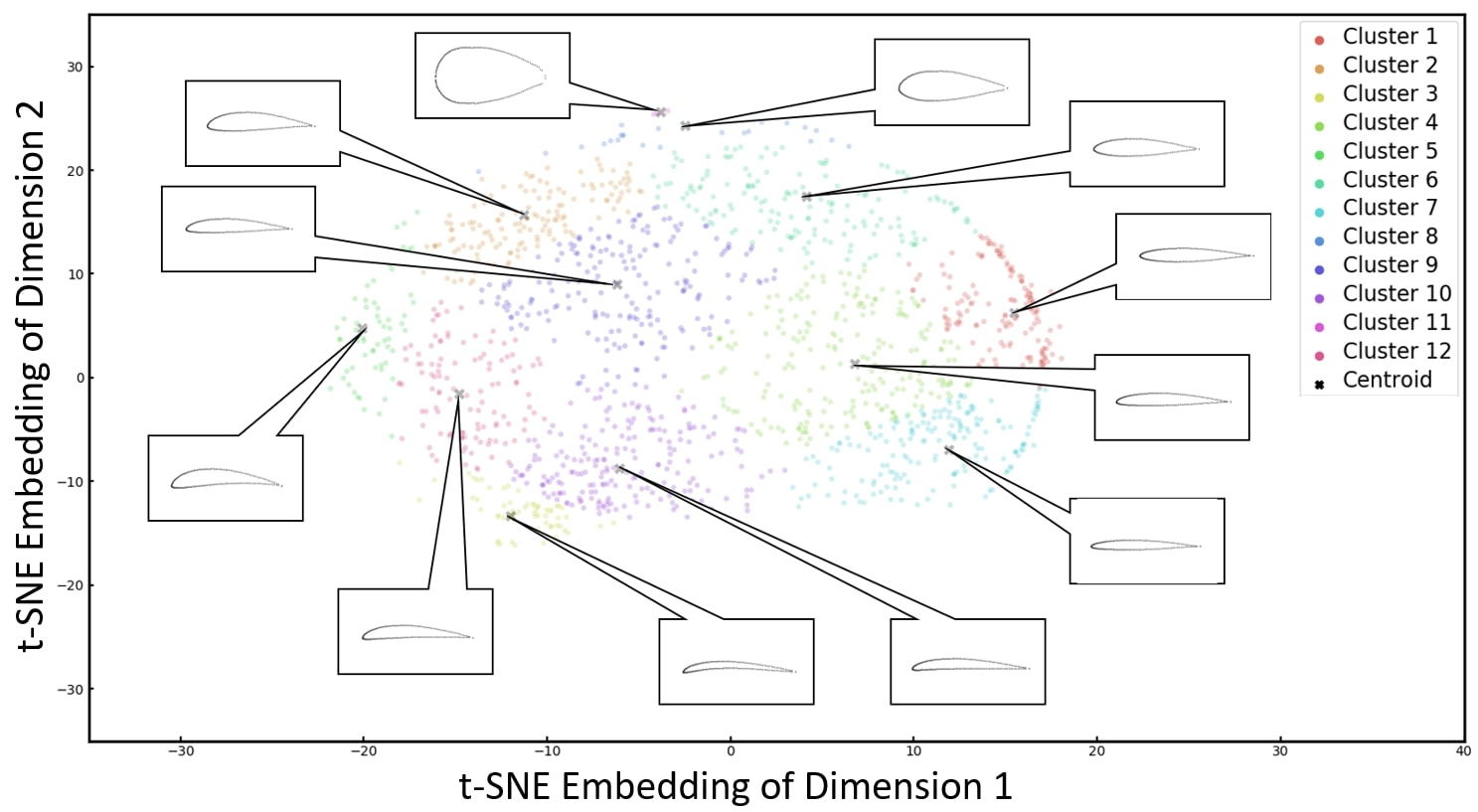}
\caption{Parametric t-SNE visualization for K-means clustering of UIUC airfoils on the learned feature domain.}
\label{fig:cluster}
\end{figure}

\noindent
The encoded features obtained from the encoder in our method can help better understand the shape of current airfoils. All airfoils from the UIUC Coordinate Database are first mapped to feature vectors, and an unsupervised learning algorithm, K-Means (\cite{jain2010kmeans}), is used to cluster these airfoils in the feature domain. To visualize the 32-dimensional feature domain, we use Parametric t-distributed Stochastic Neighbor Embedding (parametric t-SNE) (\cite{maaten2008tsne, van2009ptsne}) as a visualization tool. Parametric t-SNE is modeled by MLP, which maps the high dimensional feature vector $z_i$ into a low-dimension embedding $y_i$, while keeps the similarity between points. It converts similarities between data points to joint probabilities and minimizing the KL divergence between the joint probabilities of embedding $y_i$ and the original feature vector $z_i$. Eq.~\ref{eq:tsne} shows the cost function $C$, which t-SNE is expected to minimize: 
\begin{equation}
\begin{aligned}
    p_{j|i} = \frac{\exp(-\|z_i-z_j\|^2/2\sigma_i^2)}{\sum_{k \neq i} \exp (-\|z_i-z_k\|^2/2\sigma_i^2) }, & \;
    q_{j|i} = \frac{\exp(-\|y_i-y_j\|^2/2\sigma_i^2)}{\sum_{k \neq i} \exp (-\|y_i-y_k\|^2/2\sigma_i^2) }, \\
    C = KL(P || Q) = & \sum_i \sum_j p_{j|i} \log \frac{p_{j|i}}{q_{j|i}}, 
\end{aligned}
\label{eq:tsne}
\end{equation}
where $\sigma_i$ is calculated by a binary search given a fixed perplexity that is specified by the user (\cite{maaten2008tsne}). Fig.~\ref{fig:cluster} shows the K-means clustering results visualized with parametric t-SNE, where different colors represent different clusters, and the centroid of each cluster is also shown. In our case, we set the number of clusters to 12 as it balances the diversity and replication of airfoil shapes from different clusters for better visualization. Centroid airfoils from adjacent clusters are similar with each other, for instance, airfoils from cluster 3 and 10 are both slim and have concave bottom curves. As the distance increases in the latent feature space, the difference between airfoil shapes accumulates in different aspects, including symmetry, height, camber, etc. As the relative similarities between data are maintained via parametric t-SNE, the distance of points in Fig.~\ref{fig:cluster} reflects the distance of feature features. It is also observed that the distance of points reflects the difference in airfoil shapes, which visualizes that latent features learned by VAEGAN encode airfoil shapes.

\subsection{Learned Feature Domain Visualization}
\begin{figure}[bt!]
\centering
\includegraphics[width=0.5\textwidth]{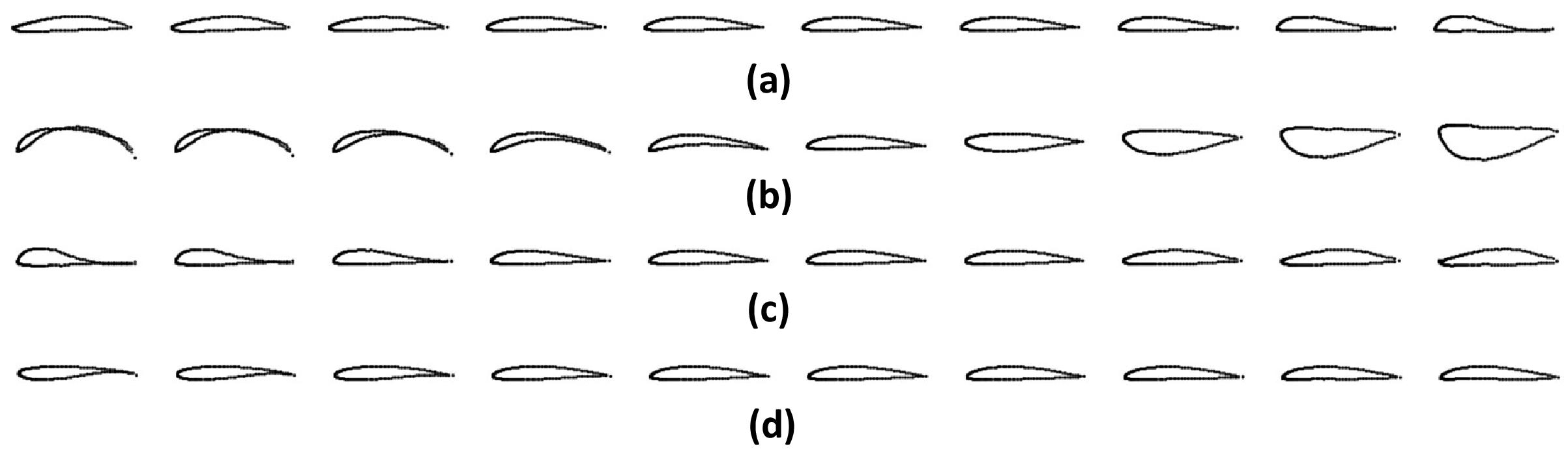}
\caption{VAEGAN-generated airfoils by gradually changing only (a) the 1\textsuperscript{st}, (b) the 8\textsuperscript{th}, (c) the 22\textsuperscript{nd}, (d) the 32\textsuperscript{nd} latent dimension.}
\label{fig:change}
\end{figure}

\begin{figure}[bt!]
\centering
\includegraphics[width=0.5\textwidth]{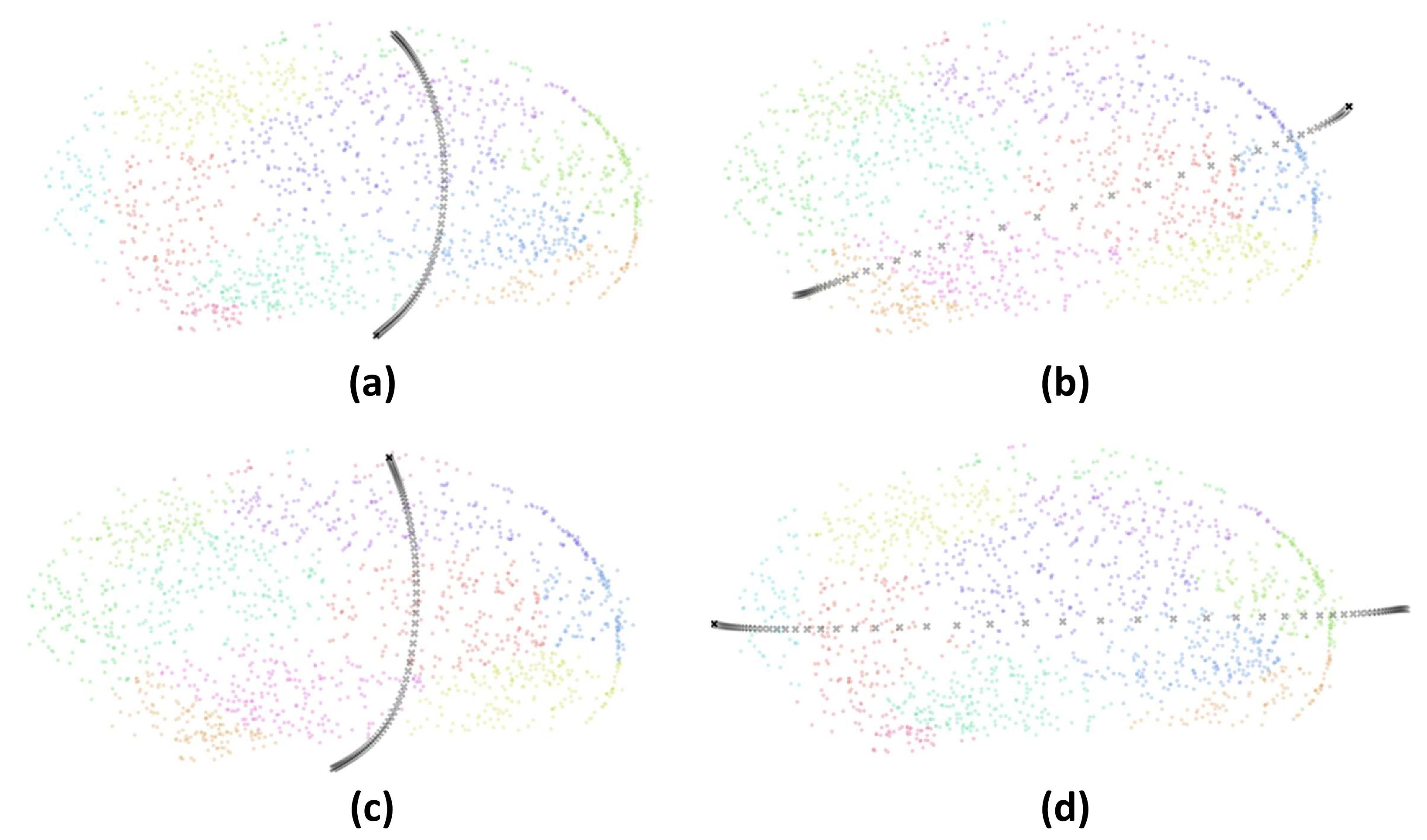}
\caption{Path visualization of gradually changing (a) the 1\textsuperscript{st}, (b) the 8\textsuperscript{th}, (c) the 22\textsuperscript{nd}, (d) the 32\textsuperscript{nd} latent dimension feature embedded by parametric t-SNE. }
\label{fig:change_cluster}
\end{figure}

\noindent
Also, experiments are conducted to investigate what geometric features are encoded in each dimension of the learned representation. A series of manually designed feature vectors are fed into the decoder, where all the elements are set to zero except for one specific dimension. That particular element is changed gradually from $-10$ to $10$, and the designed feature vectors are mapped to the airfoil coordinate domain by the decoder. Changes of generated airfoils are illustrated in Fig~\ref{fig:change}, and the 2D embedding of feature vectors using parametric t-SNE is shown in Fig~\ref{fig:change_cluster}. Here, only four dimensions are chosen for analysis purposes. 

The 1\textsuperscript{st} dimension encodes the height of the upper boundary, as illustrated in Fig~\ref{fig:change}a. As the 1\textsuperscript{st} dimension increases from $-10$ to $10$, the height of the front half airfoil increases while the tail becomes thinner. The 8\textsuperscript{th} dimension encodes the camber of both the upper boundary and the lower boundary. It is shown that by tuning the 8\textsuperscript{th} dimension, the upper boundary of the airfoil changes from a concave curve to a horizontal straight line, while the lower boundary evolves from a concave to a convex curve. Interestingly, the 22\textsuperscript{nd} dimension encodes quite similar representations as the 1\textsuperscript{st} dimension while in the opposite direction. In other words, generated airfoils from feature vectors whose 22\textsuperscript{nd} dimension change from $-10$ to $10$ are like those with the 1\textsuperscript{st} dimension change from $10$ to $-10$ as illustrated in Fig~\ref{fig:change_cluster}a and Fig~\ref{fig:change_cluster}c. Besides, Fig~\ref{fig:change}d shows how the last dimension is connected to the camber of the lower boundary. In detail, the curvature of the lower boundary decreases as the 32\textsuperscript{nd} dimension increases. 

\begin{figure}[bt!]
\centering
\includegraphics[width=0.5\textwidth]{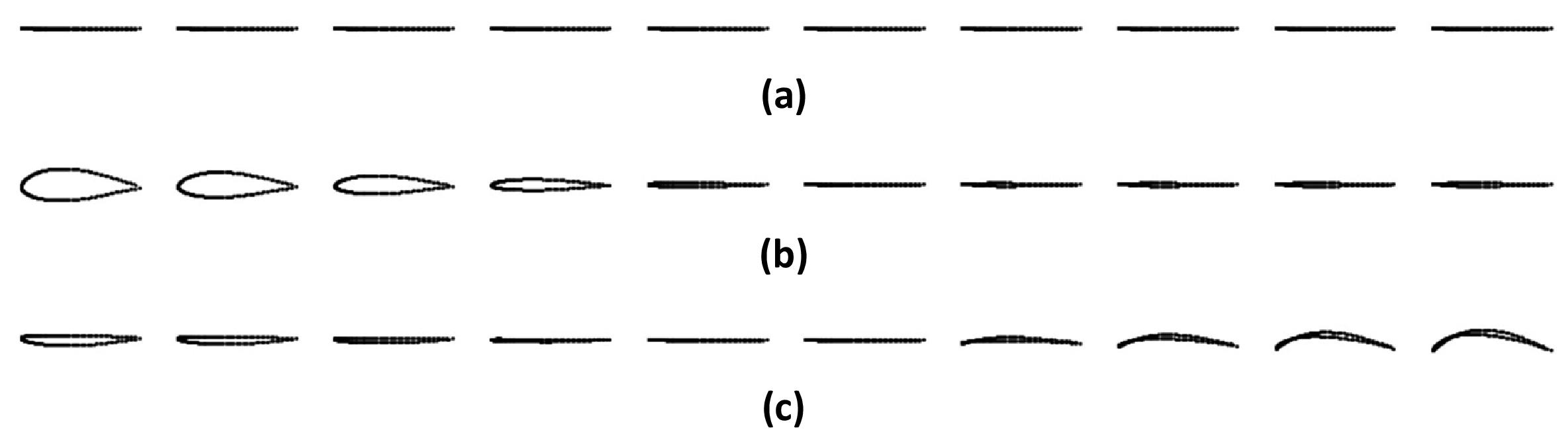}
\caption{VAE-generated airfoils by gradually changing only (a) the 1\textsuperscript{st}, (b) the 3\textsuperscript{rd}, (c) the 31\textsuperscript{st} latent dimension.}
\label{fig:vae_change}
\end{figure}

In comparison to the features learned by our VAEGAN-based model, Fig.~\ref{fig:vae_change} shows the VAE-synthesized airfoils when changing only one feature dimension. As shown in ~\ref{fig:vae_change}a, the first dimension of VAE features fails to encode any shape representations. Even in the dimensions where representations are learned as dimension 3 and 31 shown in Fig.~\ref{fig:vae_change}b and Fig.~\ref{fig:vae_change}c, the feature does not change continuously as we observe in the VAEGAN results. Our VAEGAN-based model learns more representative and thorough features than the VAE model. Also, the representations in each dimension are entangled, like the 3\textsuperscript{rd} dimension encodes both the upper bound and lower bound. 

These results indicate that, without manually designed parameters, our VAEGAN-based model learns geometrically meaningful features, and each dimension of the learned feature domain encodes informative and different geometry features. Also, each learned latent dimension can be leveraged as a design parameter when synthesizing new airfoils. The geometric information in each dimension provides new guidance for design parameter selection in comparison to conventional design methods like NACA. 

\subsection{Synthesizing Novel Airfoils}
\noindent
To make use of the encoder-decoder architecture to synthesize novel airfoils, we conduct experiments on interpolation and extrapolation of feature vectors obtained from the UIUC Database airfoils as well as sampling from random Gaussian noises following the method introduced in Section 3.3. 

\begin{figure}[bt!]
\centering
\includegraphics[width=0.5\textwidth]{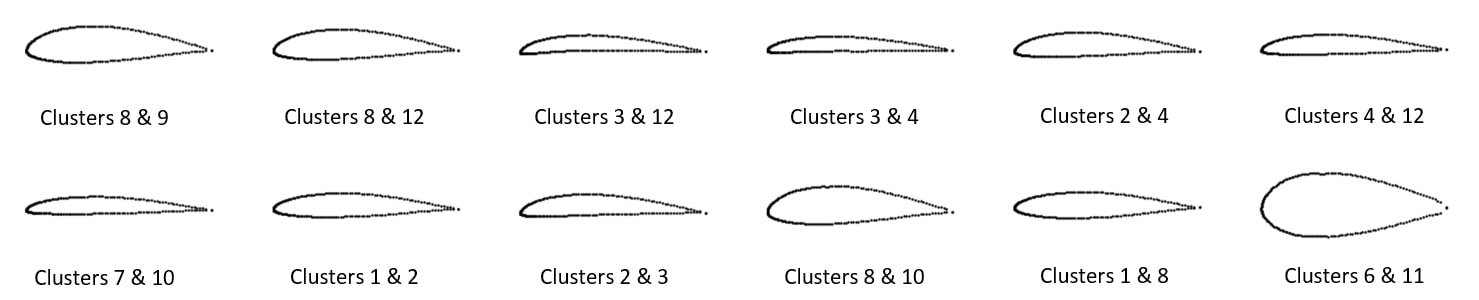}
\caption{Airfoils synthesized by interpolation of airfoils from different clusters on the latent domain.}
\label{fig:inter}
\end{figure}

\begin{figure}[bt!]
\centering
\includegraphics[width=0.5\textwidth]{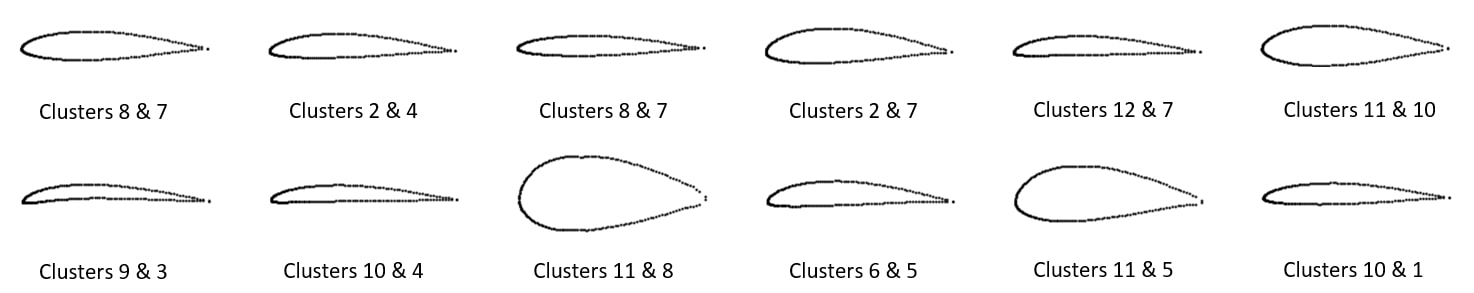}
\caption{Airfoils synthesized by extrapolation of airfoils from different clusters on the latent domain.}
\label{fig:extra}
\end{figure}

As given in Eq.~\ref{eq:inter_2}, the affine combination of two feature vectors, $z_1$ and $z_2$, are computed with $\nu=0.5$. Fig~\ref{fig:inter} shows the interpolated airfoils from two different clusters. The labels under each airfoil indicate which two clusters are $z_1$ and $z_2$ come from. As illustrated in Fig~\ref{fig:inter}, the interpolated airfoil inherits features from both clusters. For instance, Cluster 6 and Cluster 11 both represent symmetric airfoil but with variant heights. The interpolation between these two clusters synthesizes a symmetric airfoil with a medium height, as shown in the last airfoil of Fig~\ref{fig:inter}. Also, Clusters 3 and 4 both encode thin airfoils. However, the lower boundary is concave in Cluster 3, while Cluster 4 represents a convex lower boundary making the airfoil symmetric in shape. The interpolation between these two generates a thin airfoil with a flat lower boundary, which is a combination of concave and convex curves. By interpolation, novel airfoils with features from different clusters can be generated. Extrapolation between airfoils from different clusters is conducted as well. Following Eq.~\ref{eq:inter_2}, two feature vectors, $z_1$ and $z_2$, are encoded from two different airfoils, and coefficient $\nu$ is set to be $2$. Fig~\ref{fig:extra} shows the generated results from the extrapolation. Similar to interpolation, extrapolated airfoils inherit features from $z_1$ and $z_2$. 

\begin{figure*}
\centering
\includegraphics[width=.9\textwidth]{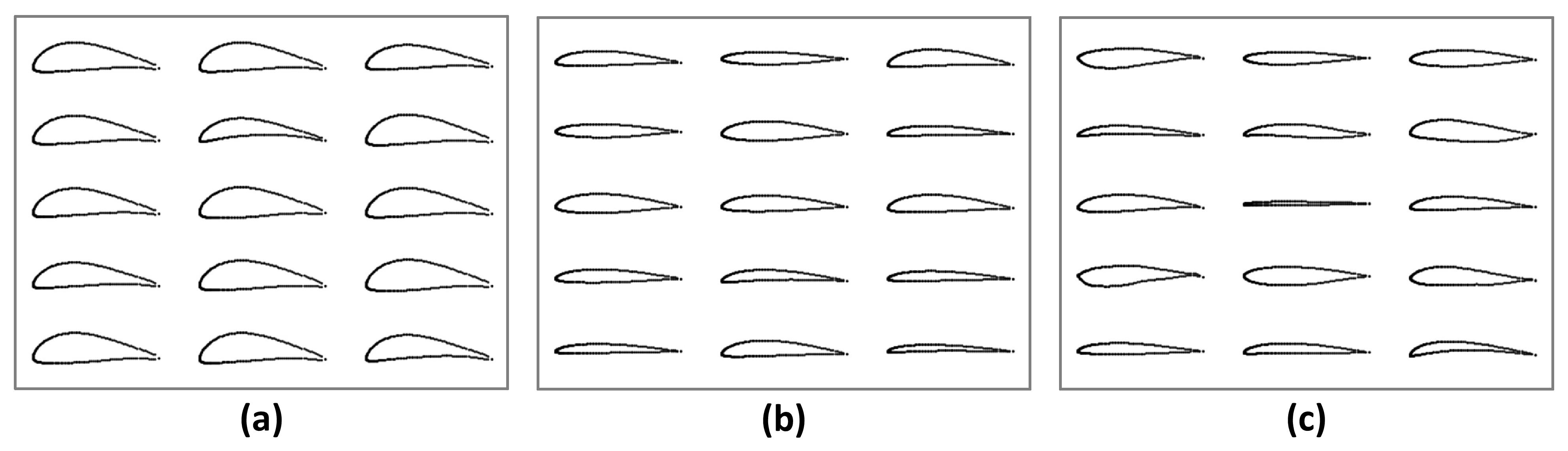}
\caption{Airfoils synthesized by sampling on the feature domain of (a) PCA, (b) VAE, (c) VAEGAN.}
\label{fig:sample}
\end{figure*}

Besides interpolation and extrapolation, the performance of the sampling synthesis method is also estimated. In sampling, a Gaussian noise, $\hat{z} \sim \mathcal{N}(0,\mathbf{I})$, is directly fed into the decoder to generate novel airfoils. Shown in Fig.~\ref{fig:sample}c are airfoils synthesized by our VAEGAN model through sampling. Besides, sampled airfoils using PCA and VAE are also included in Fig.~\ref{fig:sample}a and Fig.~\ref{fig:sample}b. 
In our implementation, we set the number of principal components as 32 in consistence with the latent dimension of the deep generative models. And each component of value $\sigma$ is sampled from a Gaussian distribution $\mathcal{N}(\sigma, 0.2 |\sigma|)$. PCA, as a linear project technique, generates similar airfoil designs through sampling, which greatly restricts the ability to guide novel designs. On the other hand, both deep-learning-based generative models, VAE and VAEGAN, can synthesize different smooth airfoil shapes. To quantitatively measure the synthesized airfoils from VAE and VAEGAN, we here introduce Fréchet Inception Distance (FID) (\cite{heusel2017gans}), which is used to evaluate the quality of samples from deep learning-based generative models. FID is calculated by computing the Fréchet distance between two feature representations. Generally, lower FID indicates higher generative sample quality. In our case, we feed the synthesized airfoils and the UIUC airfoils into the well-trained discriminator from VAEGAN and extract the second hidden layer as the representation. The FID for VAE and VAEGAN are 1.38788 and 0.65366, respectively, meaning VAEGAN synthesizes more realistic airfoils. Also, airfoils synthesized via VAEGAN possess more novelty while maintains the general geometric pattern of airfoils. For instance, in Fig.~\ref{fig:sample}c, the first airfoils in the fourth row and the third one in the second row are different from existing samples in the UIUC database. Though such novelty does not guarantee better aerodynamic properties, some airfoils are likely to have negative lift coefficients, which are infeasible in practice. The VAEGAN-based model can synthesize a wide variety of airfoils that serve as candidates for further optimization through CFD simulation as we will investigate in Section 4.6. 

\subsection{Aerodynamic Properties of Synthesized Airfoil}

\begin{figure}[bt!]
\centering
\includegraphics[width=0.48\textwidth]{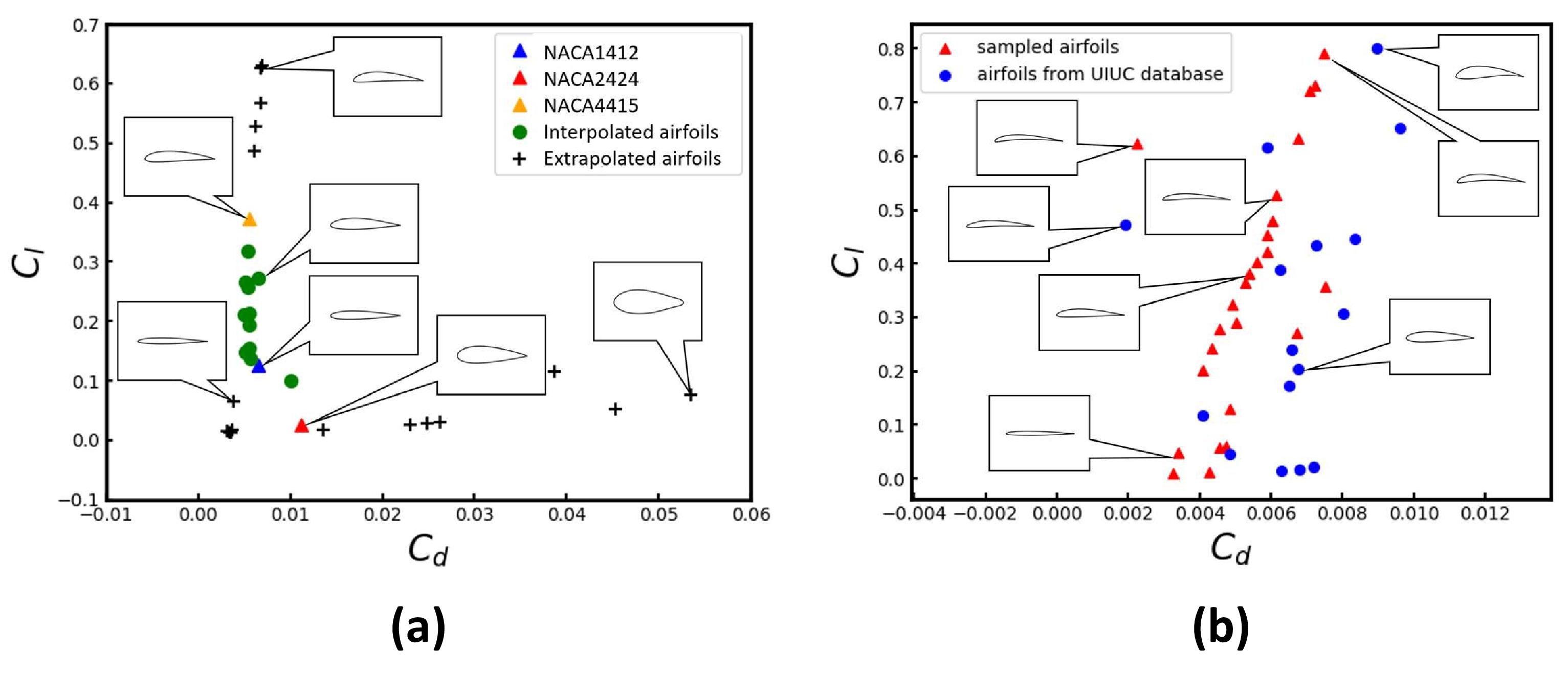}
\caption{Lift coefficient $C_l$ v.s. drag coefficient $C_d$ of (a) interpolated/extrapolated airfoils, (b) sampled airfoils.}
\label{fig:aerodyn}
\end{figure}

\noindent
Aerodynamic properties of airfoils are essential in airfoil design. Synthesized airfoils are expected to meet certain aerodynamic properties to guarantee the designs are feasible and effective. To evaluate the aerodynamics, XFoil\footnote{\url{https://web.mit.edu/drela/Public/web/xfoil/}} is utilized to compute the lift coefficient, $C_l$, and the drag coefficient, $C_d$. The experiments on XFoil are set for a low-speed condition: Reynolds number $Re=2 \times 10^6$, Mach number $Ma=0.02$, and attack angle $\alpha=\ang{0}$. As illustrated in Fig~\ref{fig:aerodyn}a, $C_l$ and $C_d$ are tested on three airfoils: NACA1412, NACA2424 and NACA4415 from the UIUC database. Following Eq.~\ref{eq:inter_3}, triplet interpolation/extrapolation is conducted with feature vectors, $z_1$, $z_2$, and $z_3$, encoded from the three NACA airfoils. The interpolated airfoils, marked by green dots, possess $C_l$ and $C_d$ in between the three NACA airfoils. While extrapolated airfoils, marked by black crossings, have significantly different aerodynamic properties from the interpolated airfoils. Some airfoils synthesized by extrapolation have high $C_l$ with relatively low $C_d$, located at the upper part of Fig.~\ref{fig:aerodyn}a. This demonstrates that by interpolation/extrapolation in the feature domain, novel airfoils with promising aerodynamic properties can be synthesized. Also, aerodynamic properties of synthesized airfoils by sampling are tested in comparison with some airfoils from the UIUC database, as shown in Fig~\ref{fig:aerodyn}b. Though the generated airfoils from Gaussian noises are not guaranteed to have good aerodynamic properties, like red dots lying on the bottom left. Some promising airfoils can be synthesized, as shown on the top right, with a high lift coefficient and a low drag coefficient. By sampling, the VAEGAN-based model can synthesize airfoils with a wide variety of aerodynamic properties. Such a variety provides abundant candidates to explore in design space for airfoil shape optimization.

\subsection{Shape Optimization on Aerodynamic Properties via Genetic Algorithm}

\begin{figure}[bt!]
\centering
\includegraphics[width=0.48\textwidth]{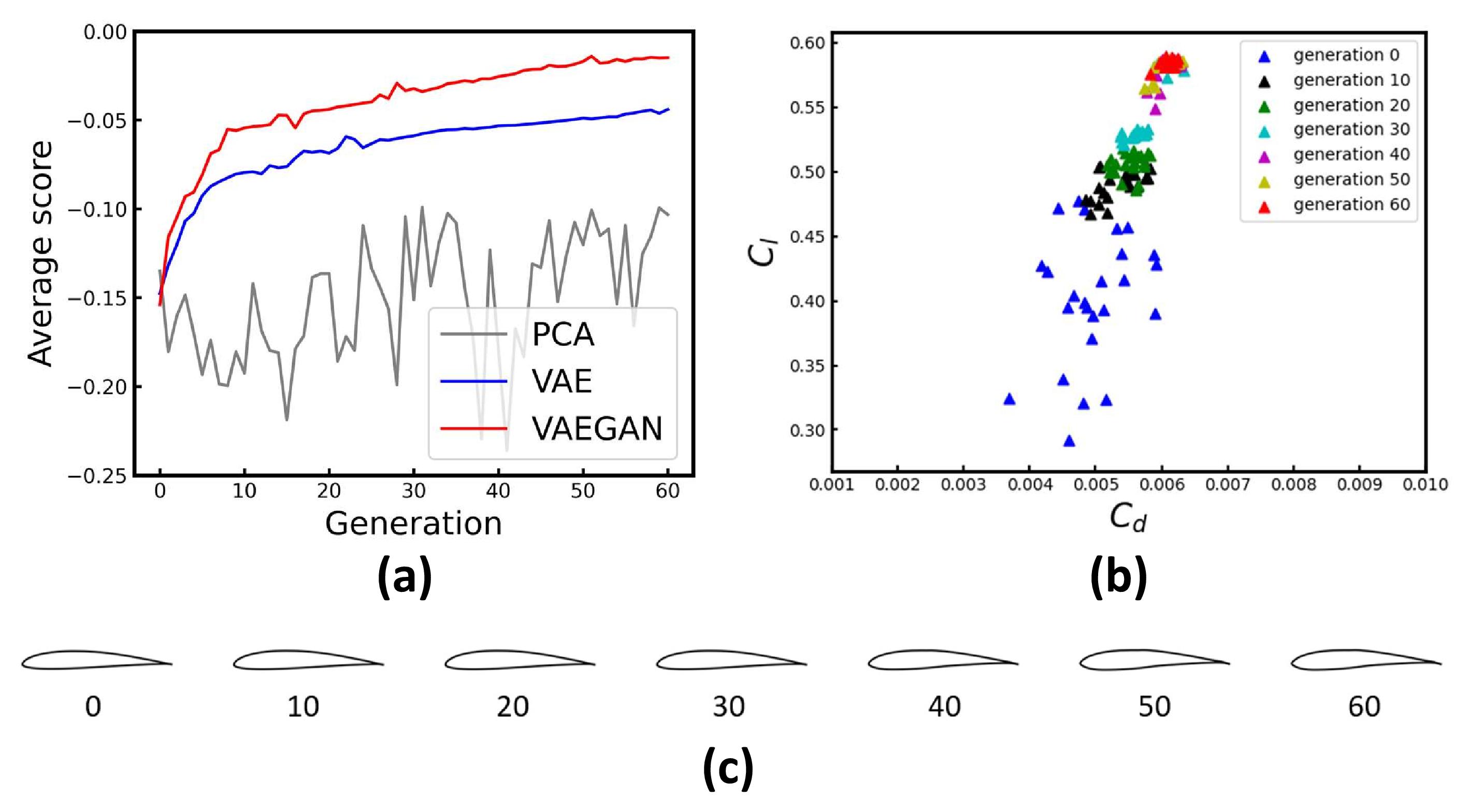}
\caption{Shape optimization via genetic algorithm: (a) average score of each generation via different generative models (i.e., PCA, VAE, and VAEGAN); (b) $C_l$ and $C_d$ of VAEGAN synthesized airfoils for different generations; (c) evolution of VAEGAN synthesized airfoils with generations.}
\label{fig:genetic_alg}
\end{figure}

\noindent
The VAEGAN model has been proven to be able to parameterize existing airfoils to latent feature vectors and synthesize novel airfoils automatically. However, whether or not the learned features and synthesized airfoils can be optimized to possess desired aerodynamic properties remains untested. To this end, this section demonstrates that with the VAEGAN model, airfoil shapes can be optimized to target $C_l^t$ and $C_d^t$ values via a genetic algorithm (GA). The lift and drag coefficients, $C_l^t$ and $C_d^t$, are calculated using XFoil under the same condition as Section 4.5. As illustrated in Fig.~\ref{fig:genetic_alg}, the target lift coefficient is $C_l^t=0.6$, and target drag coefficient is $C_d^t=0.06$. In our case, the total number of generations, $N$, is set to be 60, and the number of populations on each generation, $M$, to be 25. Fig.~\ref{fig:genetic_alg}a depicts the evolution of average score by generation when different generative methods are applied. PCA fails to generate optimized airfoils with target aerodynamic properties. On the other hand, airfoils generated from deep generative models (i.e., VAE and VAEGAN) can be optimized toward target $C_l$ and $C_d$. However, under the same number of generations and populations, VAEGAN demonstrates better performance in generating optimized airfoil designs with higher fitness scores. Fig.~\ref{fig:genetic_alg}b show how $C_l$ and $C_d$ change by generation using VAEGAN. As Fig.~\ref{fig:genetic_alg}c illustrates, the airfoil shape gradually evolves to the target $C_l$ and $C_d$. Also, we compare the performance of airfoil optimization using different featurization techniques, PCA, and VAE. The genetic algorithm with the same objective function and settings are conducted. Moreover, Table~\ref{tab:GA} lists the performance of different featurization methods when different target lift and drag coefficients $C_l^t$ and $C_d^t$ are applied. The mean and standard deviation of lift coefficients, drag coefficients, and fitness scores of the last generation are reported. For example, for $C_l^t=0.6$ and $C_d^t=0.006$, the VAEGAN-synthesized airfoils reach an average lift coefficient of 0.5857 and an average drag coefficient of 0.0061. The coefficients are close to the desired aerodynamic properties. Whereas PCA and VAE fail to synthesize desired airfoils within the same number of generations and population size. Also, it is shown that with other target aerodynamic properties, VAEGAN outperforms the other models in generating optimized airfoil designs. This is because our VAEGAN-based model generates a wider variety of airfoils that serves as potential candidates in design optimization. Such experiments prove that a simple optimization technique like GA and the well-trained VAEGAN model can synthesize airfoils with desired aerodynamic properties, which can guide the design of effective and efficient aerodynamic products. 

\begin{table*}[bt!]
\centering
\caption{Airfoil design optimization results: $C_l$, $C_d$, and fitness score, through different featurization methods for different target $C_l^t$ and $C_d^t$.}
\label{tab:GA}
\centering
\begin{tabularx}{0.9\linewidth}{cclcccccccc}
\cline{1-11}
& & & \multicolumn{2}{c}{$C_l$} & & \multicolumn{2}{c}{$C_d$} & & \multicolumn{2}{c}{Fitness score} \\
\cline{4-5} \cline{7-8} \cline{10-11}
$C_l^t$ & $C_d^t$ & Featurization & mean & std & & mean & std & & mean & std \\ 
\cline{1-11}
& & PCA    & 0.41953 & 0.012672  & & 0.0069430 & 6.9231 $\times 10^{-4}$ & & -0.12893 & 0.046250 \\
0.6 & 0.006 & VAE    & 0.52746 & 0.0017073 & & 0.0056662 & 1.4631 $\times 10^{-5}$ & & -0.017819 & 0.0014068 \\
& & VAEGAN & 0.58570 & 0.0028946 & & 0.0061030 & 7.0711 $\times 10^{-6}$  & & -8.5312 $\times 10^{-4}$ & 0.00089211 \\
\cline{1-11}
& & PCA    & 0.40342 & 0.0119782  & & 0.0066422 & 7.5323 $\times 10^{-4}$ & & -0.23431 & 0.064320 \\
0.6 & 0.005 & VAE    & 0.53245 & 0.0015981 & & 0.0055432 & 1.1176 $\times 10^{-5}$ & & -0.025632 & 0.014323  \\
& & VAEGAN & 0.57230 & 0.002061 & & 0.0053033 & 1.5986 $\times 10^{-5}$  & & -0.0057176 & 0.0012436 \\
\cline{1-11}
& & PCA    & 0.44243 & 0.024322  & & 0.0060134 & 7.4312 $\times 10^{-4}$ & & -0.056324 & 0.067883 \\
0.5 & 0.005 & VAE    & 0.49186 & 0.001168 & & 0.0053432 & 2.2679 $\times 10^{-5}$ & & -0.0051764 & 0.0012432 \\
& & VAEGAN & 0.50587 & 0.0012456 & & 0.0051879 & 5.6709 $\times 10^{-6}$  & & -0.0015501 & 0.00086793 \\
\cline{1-11}
\end{tabularx}

\end{table*}


\section{Conclusion}
\noindent
In this work, a data-driven method is proposed to achieve three goals: (1) automatically featuring airfoil geometries from the UIUC Database without manually designed parameters, (2) synthesizing novel airfoils by either interpolating or extrapolating the encoded features, as well as generating from random noise, and (3) optimizing the features to synthesize airfoils with desired aerodynamic properties. Our model is built upon VAEGAN, which combines the encoder-decoder architecture from VAE and the discriminator from GAN. With the encoder-decoder structure, our model learns explicit mappings from airfoil coordinates to latent feature domain as well as from feature vectors to airfoils, while with the discriminator, the model can automatically synthesize realistic samples. Also, our model is trained in a self-supervised manner. Namely, the model learns compact and informative features directly from airfoil shapes without manually tagged labels or designed parameters. Optimized on the learned feature domain via GA, the synthesized airfoils can evolve to have desired aerodynamic properties. 

Experiments show that our model learns compact and comprehensive features encoding shape information of airfoils and can automatically generate novel airfoils. First, airfoils can be reconstructed via decoding the learned features with minor errors compared to the origin coordinates. Second, K-Means clustering on the feature domain of the UIUC Coordinate Database further demonstrates the learned representations are meaningful in a way that the centroid of each cluster represents different shapes. It is also investigated what is encoded in each dimension of the feature domain by gradually changing the feature vector on one specific dimension with all other dimensions fixed. Without human prior, each dimension encodes different geometric information like height, camber, symmetry, and even coupled features. Moreover, novel airfoils are synthesized by interpolating and extrapolating learned features from different airfoils as well as directly generated from random noise. By interpolating or extrapolating, the synthesized airfoil inherits and blends features from existing airfoils, which provides insights for designing new airfoils. On the other hand, airfoils generated from Gaussian noise are more aggressive in a way that they follow a less geometrical format of existing airfoils, and more novelty is introduced to the airfoil design. Finally, the synthesized airfoils can be optimized via GA to possess competitive or even better aerodynamic properties in comparison to existing ones, indicating the synthesized geometries are not only plausible in shape but also practical in aerodynamic performance. 

Compared to the other featurization methods, like PCA and VAE, our model performs better in synthesizing realistic airfoil shapes and encoder more representative geometric features. Besides, our method synthesizes airfoils without pre-defined polynomials or splines formula and demonstrates deep generative models as an efficient tool for airfoil design. In the scope of this paper, we only focus on the design of 2D airfoils. Extending the deep generative models to 3D airplane wing design is of great interest to investigate. Besides, airfoil design optimization on the learned latent feature can be further investigated. For example, optimizing the airfoils under varying aerodynamic conditions. Deep reinforcement learning, which learns an optimal trajectory towards the objective, can also be implemented in optimizing the aerodynamic shape design on the latent feature domain. In UIUC database, all airfoils share the same length of 1. Including the length as a design parameter to optimize the performance of airfoils can be a valuable extension. Other optimization methods, like Bayesian optimization (\cite{snoek2012practical}), may be applied to optimize the generated airfoil designs more efficiently and reduce the computational costs of simulations to evaluate the aerodynamic performance. In future works, the proposed deep generative method could be further extended to airfoil designs under transonic or high-speed flow.


\part{Acknowledgements}

The authors thank the start-up fund provided by the Department of Mechanical Engineering at Carnegie Mellon University. 

\part{Conflict of interest statement}
The authors declared no potential conflicts of interest with respect to the research, authorship, and/or publication of this article.

\printbibliography

\end{document}